\documentclass[twocolumn,english,showpacs,preprintnumbers,amsmath,amssymb,floatfix]{revtex4}
\usepackage[T1]{fontenc}
\usepackage[latin9]{inputenc}
\usepackage{color}
\usepackage{array}
\usepackage{amstext}
\usepackage{graphicx}
\usepackage{esint}

\makeatletter

\providecommand{\tabularnewline}{\\}

\@ifundefined{textcolor}{}
{%
 \definecolor{BLACK}{gray}{0}
 \definecolor{WHITE}{gray}{1}
 \definecolor{RED}{rgb}{1,0,0}
 \definecolor{GREEN}{rgb}{0,1,0}
 \definecolor{BLUE}{rgb}{0,0,1}
 \definecolor{CYAN}{cmyk}{1,0,0,0}
 \definecolor{MAGENTA}{cmyk}{0,1,0,0}
 \definecolor{YELLOW}{cmyk}{0,0,1,0}
 }

\@ifundefined{definecolor}
 {\usepackage{color}}{}
\@ifundefined{definecolor}
 {\usepackage{color}}{}
\makeatother

\makeatother

\usepackage{babel}

\begin{document}

\title{Polarized Deeply Inelastic Scattering (DIS) Structure Functions \\
 for Nucleons and Nuclei}

\author{Ali N. Khorramian $^{a,b}$}

\email{Khorramiana@theory.ipm.ac.ir}

\author{S. Atashbar Tehrani $^{b}$}

\email{Atashbar@ipm.ir}

\author{S. Taheri Monfared $^{a,b}$}

\email{Sara.taherimonfared@gmail.com}

\author{F. Arbabifar $^{a}$}

\email{Farbabifar@gmail.com}

\author{F. I. Olness $^{c}$}

\email{olness@smu.edu}

\affiliation{$^{(a)}$ Physics Department, Semnan University, Semnan, Iran \\
 $^{(b)}$ School of Particles and Accelerators, Institute for
Research in Fundamental Sciences (IPM), P.O.Box 19395-5531, Tehran,
Iran \\
 $^{(c)}$ Department of Physics, Southern Methodist University,
Dallas, TX 75275-0175, USA }

\date{\today}
\begin{abstract}
We extract parton distribution functions (PDFs) and structure functions
from recent experimental data of polarized lepton-DIS on nucleons
at next-to-leading order (NLO) Quantum Chromodynamics. We apply the
Jacobi polynomial method to the DGLAP evolution as this is numerically
efficient. Having determined the polarized proton and neutron spin
structure, we extend this analysis to describe $^{3}$He and $^{3}$H
polarized structure functions, as well as various sum rules. We compare
our results with other analyses from the literature.
\end{abstract}


\maketitle
\tableofcontents{}

\section{Introduction}

A fundamental challenge of high energy particle physics is to understand
the spin structure of protons, neutrons, and nuclei in terms of their
parton constituents. The increasing precision of experimental data
on inclusive polarized deeply inelastic scattering (DIS) of leptons
from nucleons allows us to perform incisive QCD analyses of polarized
structure functions to reveal the spin dependent partonic structure
function of the nucleon. Polarized DIS lepton-nucleon scattering experiments
have been performed at CERN, SLAC, DESY and JLAB \citep{JLABn,CLA1pd,EMCp,E142n,HERMn,E154QCD,E154n,SMCpd,E143pd,E155d,E155p,HERMpd,COMP1},
and these processes have played a key role in our understanding of
QCD and the spin structure of the nucleon \citep{Anselmino:1994gn,Lampe:1998eu,Hughes:1999wr,filippone-01,Altarelli:2009za}.
There are several comprehensive analyses of the polarized DIS data
in the literature \citep{alt98abfr,Ball:1997sp,bou98,flo98,gor98,lea99,str99,gho00,deFlorian:2000bm,Gluck:2000dy,Bhalerao:2001rn,Leader:2001kh,Bluemlein:2002be,Goto:1999by,deFlorian:2005mw,Leader:2005ci,Bourrely:2001du,ABFR,alt97,deFlorian:2008mr,Hirai:2008aj,Blumlein:2010rn,Leader:2010dx,Blumlein:2010ri,Khorramian:2004ih,Atashbar Tehrani:2007be};
this work provides a detailed picture of the spin structure of the
nucleons.

The new precision experimental data from the HERMES and COMPASS collaborations
\citep{HERMpd,COMP1} of the spin structure function $g_{1}$ provides
additional information that we shall use to study the spin structure
and quark helicity distributions. We shall choose an approach based
on the expansion of orthogonal polynomials; specifically, we will
implement Jacobi polynomials as we use experimental data for each
bin of $Q^{2}$ separately~\citep{Khorramian:2004ih}. Previously~\citep{Atashbar Tehrani:2007be},
\textcolor{black}{we applied the Jacobi polynomials to determine the
polarized valon distributions using only the proton experimental data.
In this analysis, both the unpolarized and polarized valon distributions
were extracted, so more unknown parameters were required as compared
to the present analysis. The Jacobi polynomial expansion has also
been applied to} a variety of QCD analyses \citep{parisi,Barker,Krivokhizhin:1987rz,Krivokhizhin:1990ct,Chyla:1986eb,Barker:1980wu,Kataev:1997nc,Kataev:1998ce,Kataev:1999bp,Kataev:2001kk,Khorramian:2007zz,Khorramian:2008zz,Khorramian:2008zza,AtashbarTehrani:2009zz,Khorramian:2009zz1,Khorramian:2009zz2,Khorramian:2009zz3,Khorramian:2009zz4,Khorramian:2008yh,Khorramian:2009zz,Khanpour:2009zz,Khorramian:2009xz},
including the case of polarized PDFs \citep{arbabifar,Mirjalili:2009zz,Atashbar
Tehrani:2007be,Khorramian:2007gu,Khorramian:2007zza,Mirjalili:2007ep,Mirjalili:2006hf,Leader:1997kw}.

In the present study, we perform a NLO QCD analysis of the polarized
deep--inelastic data \citep{E143pd,HERMn,HERMpd,SMCpd,EMCp,E155p,COMP1,E142n,E154QCD,E154n,E155d}
in the $\overline{{\rm MS}}$--scheme and extract parameterizations
of the polarized PDFs and structure functions. In Section~II, we
provide an overview of the Jacobi polynomials approach. In Section~III
we review the parametrization and evolution of the PDFs. In Section~IV
we present the results of our fit to the data, and in Section~V we
compute the associated structure functions and sum rules. Section~VI
contains the conclusions. We also provide an Appendix which describes
the \texttt{FORTRAN}-code which is available.

\section{The Jacobi Polynomial Method}

We perform a NLO fit of the polarized parton distributions (PPFDs)
using Jacobi polynomials to reconstruct the $x$ dependent quantities
from their Mellin moments. The use of Jacobi polynomials has a number
of advantages; specifically, it will allow us to factorize the $x$
and $Q^{2}$ dependence in a manner that allows an efficient parameterization
and evolution of the structure functions.

For example, if we consider the spin structure function $xg_{1}(x,Q^{2})$,
we can expand this as:

\begin{equation}
xg_{1}(x,Q^{2})=x^{\beta}(1-x)^{\alpha}\ \sum_{n=0}^{N_{max}}a_{n}(Q^{2})\ \Theta_{n}^{\alpha,\beta}(x)\ .\label{eq:xg1}\end{equation}
 Here, $\Theta_{n}^{\alpha,\beta}(x)$ are Jacobi polynomials of order
$n$, and $N_{max}$ is the maximum order of our expansion. In this
instance, the Jacobi polynomials allow us to factor out the essential
part of the $x$-dependence of the structure function into a weight
function \citep{parisi}, and the $Q^{2}$-dependence is contained
in the Jacobi moments $a_{n}(Q^{2})$.

To be more specific, the $x$-dependence of the Jacobi polynomials
can be written as

\begin{equation}
\Theta_{n}^{\alpha,\beta}(x)=\sum_{j=0}^{n}c_{j}^{(n)}(\alpha,\beta)\ x^{j},\label{eq:jac}\end{equation}
 where the $c_{j}^{(n)}(\alpha,\beta)$ coefficients are combinations
of $\Gamma$-functions involving $\{n,\alpha,\beta\}$. The Jacobi
polynomials satisfy an orthogonality relation with weight function
$x^{\beta}(1-x)^{\alpha}$ as follows:

\begin{equation}
\int_{0}^{1}dx\; x^{\beta}(1-x)^{\alpha}\Theta_{k}^{\alpha,\beta}(x)\Theta_{l}^{\alpha,\beta}(x)=\delta_{k,l}\ .\label{eq:ortho}\end{equation}
 Thus, given the Jacobi moments $a_{n}(Q^{2})$, the polarized structure
function $xg_{1}(x,Q^{2})$ may be reconstructed from Eq.~(\ref{eq:xg1})
\citep{Atashbar Tehrani:2007be}.

We can compute the Jacobi moments $a_{n}(Q^{2})$ using the orthogonality
relation to invert Eq.~(\ref{eq:xg1}) to obtain: \begin{eqnarray}
a_{n}(Q^{2}) & = & \int_{0}^{1}dx\; xg_{1}(x,Q^{2})\Theta_{k}^{\alpha,\beta}(x)\nonumber \\
 & = & \sum_{j=0}^{n}c_{j}^{(n)}(\alpha,\beta)\ {\bf M}[xg_{1},j+2]~\ .\label{eq:aMom}\end{eqnarray}
 In Eq.~(\ref{eq:aMom}), we have substituted Eq.~(\ref{eq:xg1})
for $xg_{1}(x,Q^{2})$ and introduced the Mellin transform:

\begin{eqnarray}
{\bf {M}}[xg_{1},N] & \equiv & \int_{0}^{1}dx\ x^{N-2}\ xg_{1}(x,Q^{2})\ .\label{eq:Mellin}\end{eqnarray}
 We can now relate the polarized structure function $xg_{1}(x,Q^{2})$
with its moments \citep{Atashbar Tehrani:2007be}

\begin{eqnarray}
xg_{1}(x,Q^{2}) & = & x^{\beta}(1-x)^{\alpha}\sum_{n=0}^{N_{max}}\Theta_{n}^{\alpha,\beta}(x)\nonumber \\
 & \times & \sum_{j=0}^{n}c_{j}^{(n)}{(\alpha,\beta)}\ {\bf {M}}[xg_{1},{\normalcolor {\color{red}{\normalcolor j+2}}}]\ .\label{eg1Jacob}\end{eqnarray}
 Given Eq.~(\ref{eg1Jacob}) for $xg_{1}(x,Q^{2})$, we choose the
set $\{N_{max},\alpha,\beta\}$ to achieve optimal convergence of
this series throughout the kinematic region constrained by the data.
In practice, we find $N_{max}=9$, $\alpha=3.0$, and $\beta=0.5$
to be sufficient.

\section{QCD Analysis \& Parametrization}

\subsection{Parameterization}

We consider a proton comprised of massless partons with helicity distributions
$q_{\pm}(x,Q^{2})$ which carry momentum fraction $x$ with a characteristic
scale $Q$. The difference $\delta q(x,Q^{2})=q_{+}(x,Q^{2})-q_{-}(x,Q^{2})$
measures how much the parton of flavor $q$ {}``remembers\textcolor{blue}{''}
of the parent proton polarization. We will parameterize these polarized
PDFs at initial scale $Q_{0}^{2}=4$ GeV$^{2}$ using the following
form: \begin{equation}
x\:\delta q(x,Q_{0}^{2})={\cal N}_{q}\eta_{q}x^{a_{q}}(1-x)^{b_{q}}(1+c_{q}x)\ ,\label{eq:parm}\end{equation}
 where the polarized PDFs are determined by parameters $\{\eta_{q},a_{q},b_{q},c_{q}\}$,
and the generic label $q=\{u_{v},d_{v},\bar{q},g\}$ denotes the partonic
flavors up-valence, down-valence, sea, and gluon, respectively. The
normalization constants ${\cal N}_{q}$ \begin{equation}
\frac{1}{{\cal N}_{q}}=\left(1+c_{q}\frac{a_{q}}{a_{q}+b_{q}+1}\right)\, B\left(a_{q},b_{q}+1\right)\ ,\label{eq:norm}\end{equation}
 are chosen such that $\eta_{i}$ are the first moments of $\delta q_{i}(x,Q_{0}^{2})$,
$\eta_{i}=\int_{0}^{1}dx\delta q_{i}(x,Q_{0}^{2})$, where $B(a,b)$
is the Euler beta function.

The total up and down PDFs are a sum of the valence plus sea distributions:
$\delta u=\delta u_{v}+\delta\bar{q}$ and $\delta d=\delta d_{v}+\delta\bar{q}$.
We will assume an $SU(3)$ flavor symmetry such that $\delta\overline{q}\equiv\delta\overline{u}=\delta\overline{d}=\delta s=\delta\overline{s}$.
While we could allow for an $SU(3)$ symmetry violation term by introducing
$\kappa$ such that $\delta s=\delta\overline{s}=\kappa\delta\bar{q}$,
as the strange PDF is poorly constrained the results would be insensitive
to the specific choice of $\kappa$.

As seen from Eq.~(\ref{eq:parm}), each of four polarized parton
densities $q=\{u_{v},d_{v},\bar{q},g\}$ contain four parameters $\{\eta_{q},a_{q},b_{q},c_{q}\}$
which gives a total of 16 parameters that we must constrain. We now
demonstrate that we can eliminate some of these parameters while maintaining
sufficient flexibility to obtain a good fit.

\subsubsection{First Moments of $\delta u_{v}$ and $\delta d_{v}$}

The parameters $\eta_{u_{v}}$ and $\eta_{d_{v}}$ are the first moments
of the $\delta u_{v}$ and $\delta d_{v}$ polarized valence quark
densities; these quantities can be related to $F$ and $D$ as measured
in neutron and hyperon $\beta$--decays according to the relations
\citep{PDG}\textcolor{blue}{: }

\begin{eqnarray}
a_{3} & = & \int_{0}^{1}dx\:\delta q_{3}=\eta_{u_{v}}-\eta_{d_{v}}=F+D\ ,\\
a_{8} & = & \int_{0}^{1}dx\:\delta q_{8}=\eta_{u_{v}}+\eta_{d_{v}}=3F-D\ .\end{eqnarray}
 where $a_{3}$ and $a_{8}$ are non-singlet combinations of the first
moments of the polarized parton densities corresponding to \begin{eqnarray}
q_{3} & = & (\delta u+\delta\overline{u})-(\delta d+\delta\overline{d})\ ,\\
q_{8} & = & (\delta u+\delta\overline{u})+(\delta d+\delta\overline{d})-2(\delta s+\delta\overline{s})\ {\color{blue}.}\end{eqnarray}

\noindent A reanalysis of $F$ and $D$ with updated $\beta$-decay
constants obtained \citep{PDG} $F=0.464\pm0.008$ and $D=0.806\pm0.008$.
With these values we find:

\begin{eqnarray}
\eta_{u_{v}} & = & +0.928\pm0.014\ ,\label{eq:etauv}\\
\eta_{d_{v}} & = & -0.342\pm0.018\ .\label{eq:etadv}\end{eqnarray}

\noindent We make use of $\eta_{u_{v}}$ and $\eta_{d_{v}}$ to reduce
the number of parameters by two.

\subsubsection{Gluon and Sea-Quarks}

We find the factor $(1+c_{q}x)$ in Eq.~(\ref{eq:parm}) provides
flexibility to obtain a good description of the data, particularly
for the valence distributions $\{u_{v},d_{v}\}$. Thus we will make
use of the $c_{q}$ coefficients for the the up-valence and down-valence
distributions; in contrast, we are able to set the values for $c_{\bar{q}}$
and $c_{g}$ to zero $(c_{\bar{q}}=c_{g}=0)$ while maintaining a
good fit and eliminating two free parameters. For the parameters $\{c_{u_{v}},c_{d_{v}}\}$
we find the fit improves if we use non-zero values, but as these are
relatively flat directions in $\chi$-space we shall fix the values
as detailed in Table~I.

Separately, we find the $b$ parameters control the large-$x$ behavior
of the PDFs; thus, the sea-quark and gluon distributions have large
uncertainties in this region as they are dominated by the valence.
To provide some guidance, we observe that for \emph{unpolarized} parton
densities in the large-$x$ region, a ratio of $b_{\bar{q}}/b_{g}\sim1.6$
provides a good fit. Therefore we impose this ratio on the \emph{polarized}
$b_{\bar{q}}$ and $b_{g}$ parameters to further reduce the free
parameters. Additionally, we are able to extract reasonable constraints
on the $a_{\bar{q}}$ and $a_{g}$ parameters; this is a benefit of
the Jacobi polynomials.

Having fixed $\{\eta_{u_{v}},\eta_{d_{v}},c_{\bar{q}},c_{g}\}$ and
the ratio $b_{\bar{q}}/b_{g}$ in preliminary minimization, we then
set the parameters $\{b_{\bar{q}},b_{g},c_{u_{v}},c_{d_{v}}\}$ as
indicated in Table~\ref{tab:fit}; this gives us a total of 9 unknown
parameters, in addition to $\alpha_{s}(Q_{0}^{2})$.

\subsection{DGLAP Evolution }

In the Jacobi polynomial approach the DGLAP evolution equations are
solved in Mellin space. The Mellin transform of the parton densities
$q$ are defined analogous to that of Eq.~(\ref{eq:Mellin}): \begin{eqnarray}
\textbf{M}[\delta q(x,Q_{0}^{2}),N] & \equiv & \delta q(N,Q_{0}^{2})=\int_{0}^{1}x^{N-1}\:\delta q(x,Q_{0}^{2})\: dx\nonumber \\
 & = & {\cal {N}}_{q}\eta_{q}\left(1+c_{q}\:\frac{N-1+a_{q}}{N+a_{q}+b_{q}}\right)\ \nonumber \\
 & \times & B(N-1+a_{q},b_{q}+1)\ ,\end{eqnarray}
where $q=\{u_{v},d_{v},\overline{q},g\}$, and $B$ is the Euler beta
function.

In Mellin space, the twist-2 contributions to the polarized structure
function $g_{1}(N,Q^{2})$ can be represented in terms of the polarized
parton densities and the coefficient functions $\Delta C_{i}^{N}$
by:\begin{eqnarray}
\textbf{M}[g_{1}^{p},N] & = & \frac{1}{2}\sum\limits _{q}e_{q}^{2}\left\{ (1+\frac{\alpha_{s}}{2\pi}\Delta C_{q}^{N})\right.\nonumber \\
 & \times & [\delta q(N,Q^{2})+\delta\bar{q}(N,Q^{2})]\nonumber \\
 & + & \left.\frac{\alpha_{s}}{2\pi}\:2\Delta C_{g}^{N}\delta g(N,Q^{2})\right\} \;.\end{eqnarray}
 Here, the sum runs over quark flavors $\{u,d,s\}$, and $\{\delta q,\delta\bar{q},\delta g\}$
are the polarized quark, anti-quark, and gluon distributions, respectively.

The coefficient functions $\Delta C_{i}^{N}$ are the $N$-th moments
of spin-dependent Wilson coefficients, and are given by \citep{Lampe:1998eu}:

\begin{eqnarray*}
\Delta C_{q}^{N} & = & \frac{4}{3}\:\left\{ -S_{2}(N)+(S_{1}(N))^{2}+\left(\frac{3}{2}-\frac{1}{N(N+1)}\right)\right.\\
 & \times & \left.S_{1}(N)+\frac{1}{N^{2}}+\frac{1}{2N}+\frac{1}{N+1}-\frac{9}{2}\right\} ,\end{eqnarray*}
 \begin{eqnarray*}
\Delta C_{g}^{N} & = & \frac{1}{2}\left[-\frac{N-1}{N(N+1)}(S_{1}(N)+1)-\frac{1}{N^{2}}+\frac{2}{N(N+1)}\right]\ ,\end{eqnarray*}
 with $S{}_{1}(n)=\sum_{j=1}^{n}\frac{1}{j}=\psi(n+1)+\gamma_{E}$,
$S{}_{2}(n)=\sum_{j=1}^{n}\frac{1}{j^{2}}=(\frac{\pi^{2}}{6})-\psi'(n+1)$,
$\psi(n)=\Gamma'(n)/\Gamma(n)$ and $\psi'(n)=d^{2}\ln\Gamma(n)/dn^{2}$.

In summary, we are able to express $xg_{1}^{p}$ in terms of 9 unknown
parameters at an input scale of ${\normalcolor Q_{0}^{{\color{blue}{\normalcolor 2}}}=4}$
GeV$^{2}$. We now examine the fits to the spin structure functions
to extract the polarized PDFs from the available data.

\section{QCD fit of ${\bf xg_{1}}(x,Q^{2})$ data}

\global\long\def\fstrut{\rule{0pt}{12pt}}
\begin{table}
\begin{tabular}{>{\centering}p{0.3in}>{\centering}p{0.3in}c>{\centering}p{0.3in}>{\centering}p{0.3in}c}
\hline
 & $\eta_{u_{v}}$  & $~0.928\ (fixed)~$  &  & $\eta_{\bar{q}}$  & $-0.054\pm0.029$ \tabularnewline
$\delta u_{v}$  & $a_{u_{v}}$  & $0.535\pm0.022$  & $\delta\bar{q}$  & $a_{\bar{q}}$  & $0.474\pm0.121$ \tabularnewline
 & $b_{u_{v}}$  & $3.222\pm0.085$  &  & $b_{\bar{q}}$  & $9.310\ (fixed)$ \tabularnewline
 & $c_{u_{v}}$  & $8.180\ (fixed)$  &  & $c_{\bar{q}}$  & $0$ \tabularnewline
\hline
 & $\eta_{d_{v}}$  & $-0.342\ (fixed)$  &  & $\eta_{g}$  & $0.224\pm0.118$ \tabularnewline
$\delta d_{v}$  & $a_{d_{v}}$  & $0.530\pm0.067$  & $\delta g$  & $a_{g}$  & $2.833\pm0.528$ \tabularnewline
 & $b_{d_{v}}$  & $3.878\pm0.451$  &  & $b_{g}$  & $5.747\ (fixed)$ \tabularnewline
 & $c_{d_{v}}$  & $~4.789\ (fixed)~$  &  & $c_{g}$  & $0$ \tabularnewline
\hline
\multicolumn{6}{c}{\fstrut $\alpha_{s}(Q_{0}^{2})\ =\ 0.381\pm0.017$}\tabularnewline
\hline
\multicolumn{6}{c}{\fstrut $\chi^{2}/dof\ =\ 273.6/370\ =\ 0.74$}\tabularnewline
\hline
\end{tabular}

\caption{{\small Final parameter values and their statistical errors in the
$\overline{{\rm MS}}$--scheme at the input scale $Q_{0}^{2}=4$ GeV$^{2}$.}
\label{tab:fit}}

\end{table}

\begin{table*}
\begin{tabular}{ccccc}
\hline
\textbf{Experiment}  & \textbf{$x$-range}  & \textbf{Q$^{2}$-range{[}GeV$^{2}${]}}  & \textbf{\# of data points}  & \textbf{${\cal {N}}_{i}$} \tabularnewline
\hline
\hline
E143 (p)  & 0.031-0.749  & 1.27-9.52  & 28  & 0.9998 \tabularnewline
HERMES \emph{(p)}  & 0.028-0.66  & 1.01-7.36  & 39  & 1.0006 \tabularnewline
SMC \emph{(p)}  & 0.005-0.480  & 1.30-58.0  & 12  & 0.9999 \tabularnewline
EMC \emph{(p)}  & 0.015-0.466  & 3.50-29.5  & 10  & 1.0094 \tabularnewline
E155  & 0.015-0.750  & 1.22-34.72  & 24  & 1.0226 \tabularnewline
HERMES06 \emph{(p)}  & 0.026-0.731  & 1.12-14.29  & 51  & 0.9992 \tabularnewline
COMPASS10 \emph{(p)}  & 0.005-0.568  & 1.10-62.10  & 15  & 0.9920 \tabularnewline
\textbf{Proton}  &  &  & \textbf{179}  & \tabularnewline
\hline
E143 \emph{(d)}  & 0.031-0.749  & 1.27-9.52  & 28  & 0.9990 \tabularnewline
E155 \emph{(d)}  & 0.015-0.750  & 1.22-34.79  & 24  & 0.9998 \tabularnewline
SMC \emph{(d)}  & 0.005-0.479  & 1.30-54.80  & 12  & 0.9999 \tabularnewline
HERMES06 \emph{(d)}  & 0.026-0.731  & 1.12-14.29  & 51  & 0.9976 \tabularnewline
\textbf{Deuteron}  &  &  & \textbf{115}  & \tabularnewline
\hline
E142 \emph{(n)}  & 0.035-0.466  & 1.10-5.50  & 8  & 0.9991 \tabularnewline
HERMES \emph{(n)}  & 0.033-0.464  & 1.22-5.25  & 9  & 0.9999 \tabularnewline
E154 \emph{(n)}  & 0.017-0.564  & 1.20-15.00  & 17  & 0.9996 \tabularnewline
HERMES06 \emph{(n)}  & 0.026-0.731  & 1.12-14.29  & 51  & 1.0000 \tabularnewline
\textbf{Neutron}  &  &  & \textbf{85}  & \tabularnewline
\hline
\hline
\textbf{Total}  &  &  & \textbf{379}  & \tabularnewline
\hline
\end{tabular}

\caption{{\small Published data points with the measured $x$ and $Q^{2}$
ranges, the number of data points (with a cut of $Q^{2}\geq1.0$ GeV$^{2}$),
and the fitted normalization shifts ${\cal {N}}_{i}$. \label{tab:data}}}

\end{table*}

Our analysis is performed using the QCD-PEGASUS program \citep{Vogt:2004ns}.
We work at NLO in the QCD evolution using $N_{f}=3$ in the fixed-flavor
number scheme with massless partonic flavors $\{u,d,s\}$. We take
the renormalization and factorization scales to be equal $(\mu_{R}=\mu_{F})$,
and we compute the strong coupling $a_{s}(Q^{2})$ at NLO using a
fourth order Runge-Kutta integration. Our initial parameterizations
(Eq.~\ref{eq:parm}) are chosen to be invertible in N-space, and
this makes our fitting procedure numerically efficient.

For the proton data we use EMC \citep{EMCp}, HERMES \citep{HERMn,HERMpd},
SMC \citep{SMCpd}, E143 \citep{E143pd}, E155 \citep{E155p} and
COMPASS \citep{COMP1}, for the neutron data we use E142\citep{E142n},
HERMES \citep{HERMn,HERMpd} and E154 \citep{E154QCD,E154n}, and
for the deuteron data we use SMC \citep{SMCpd}, E143 \citep{E143pd},
E155 \citep{E155d} and HERMES\citep{HERMpd}. This data is summarized
in Table~\ref{tab:data}.

We minimize the global $\chi^{2}$\citep{Stump:2001gu,Khorramian:2008yh,Khorramian:2009xz}:

\begin{equation}
\chi_{\mathrm{global}}^{2}=\sum_{n}w_{n}\chi_{n}^{2}\;,\label{eq:chi2}\end{equation}
 where the sum $n$ runs over the different experiments, $w_{n}$
is a weight factor for the $n$-th experiment, and $\chi_{n}^{2}$
is given by:

\begin{equation}
\chi_{n}^{2}=\left(\frac{1-{\cal N}_{n}}{\Delta{\cal N}_{n}}\right)^{2}+\sum_{i}\left(\frac{{\cal N}_{n}\: g_{1,i}^{exp}-g_{1,i}^{theor}}{{\cal N}_{n}\:\Delta g_{1,i}^{exp}}\right)^{2}\;.\end{equation}
 Here, $g_{1,i}^{exp}$, $\Delta g_{1,i}^{exp}$, and $g_{1,i}^{theor}$
denote the experimental measurement, the experimental uncertainty
(statistical and systematic combined in quadrature) and theoretical
value for the $i^{\mathrm{th}}$ data point, respectively. ${\Delta{\cal N}_{n}}$
is the experimental normalization uncertainty and ${\cal N}_{n}$
is an overall normalization factor for the data of experiment $n$.
We allow for a relative normalization shift ${\cal N}_{n}$ between
different data sets within uncertainties ${\Delta{\cal N}_{n}}$ quoted
by the experiments.

We minimize the above $\chi^{2}$ value with the 9 unknown parameters
plus an undetermined $\alpha_{s}(Q_{0}^{2})$. The values of these
parameters are summarized in Table~\ref{tab:fit}. We find ${\chi}^{2}/{\rm {d.o.f.}}=273.6/370$
which yields an acceptable fit to the experimental data.

\section{PDF and Structure Function Analysis}

We next present our polarized PDFs and perform comparisons with other
recent parameterizations \citep{Bluemlein:2002be,Gluck:2000dy,Leader:2005ci,deFlorian:2005mw,Goto:1999by}.

\subsection{Polarized PDFs }

\begin{figure}
\includegraphics[clip,width=0.48\textwidth]{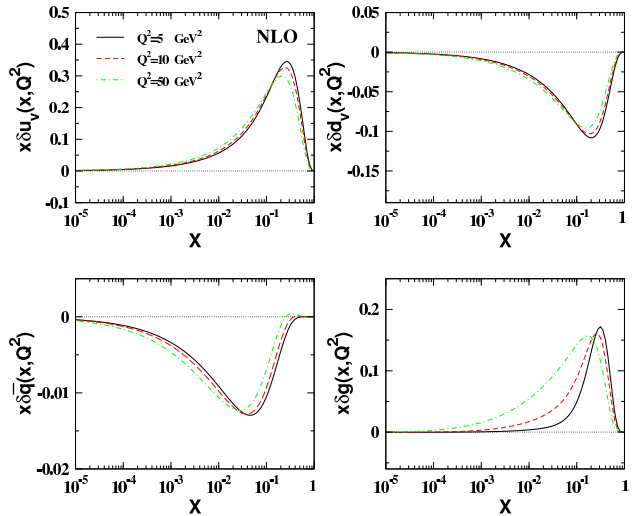}

\caption{{\small The polarized parton distribution as function of $x$ and
for different values of $Q^{2}$. \label{fig:PPDFS}}}

\end{figure}

\begin{figure}
\includegraphics[clip,width=0.48\textwidth]{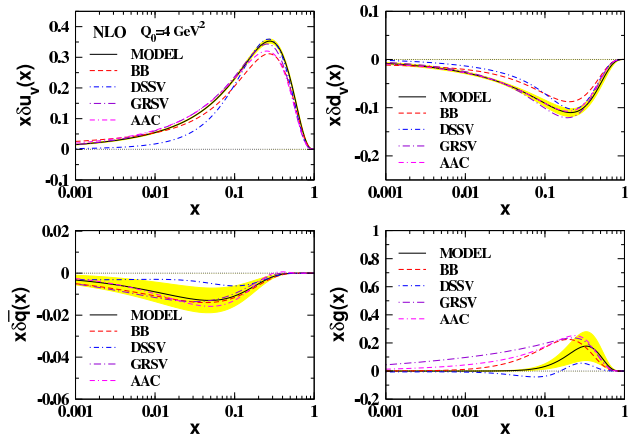}

\caption{{\small The polarized parton distribution at $Q_{0}^{2}=$ 4 GeV$^{2}$
as a function of $x$. Our fit is the solid curve. Also shown are
the results of BB (dashed) \citep{Blumlein:2010rn}, DSSV (dashed-dotted)
\citep{deFlorian:2008mr}, GRSV (long dashed-dotted) \citep{Gluck:2000dy},
and AAC (dashed-dashed-dotted) \citep{Hirai:2008aj}. \label{fig:PDFcompare}}}

\end{figure}

Figure~\ref{fig:PPDFS} displays our polarized PDFs for a selection
of $Q^{2}$ values. The up-valence ($x\delta u_{v}$) and gluon ($x\delta g$)
distributions are positive, while the down-valence ($x\delta d_{v}$)
and sea ($x\delta\bar{q}$) distributions are negative. We observe
that the evolution shifts all the distributions to smaller values
of $x$, and tends to flatten out the peak for increasing $Q^{2}$.
Figure~\ref{fig:PDFcompare} displays the extracted NLO polarized
PDFs as compared with various parameterizations from the literature
\citep{Blumlein:2010rn,deFlorian:2008mr,Gluck:2000dy,Hirai:2008aj}.

\textcolor{black}{Examining the $x\delta u_{v}$ and $x\delta\bar{q}$
distributions we see that most of the fits are in agreement, with
the possible exception of the DSSV \citep{deFlorian:2008mr} curves;
for both distributions, the DSSV results approach zero more quickly
than the other curves. For the $x\delta d_{v}$ distribution, all
of the curves are comparable. The DSSV analysis employs results from
semi-inclusive DIS (SI-DIS) data which can impose individual constraints
on individual quark flavor distributions in the nucleon \citep{deFlorian:2008mr}.
Finally, for the gluon distribution, the DSSV results have a sign
change in the region of $x\sim0.1$ while the other fits are positive.
Our result for gluon distribution is located between DSSV curve and
the other fits \citep{Blumlein:2010rn,Gluck:2000dy,Hirai:2008aj}.
In particular, we find the gluon polarization vanished more quickly
for small $x$ values as compared with the other fits;}\textcolor{black}{\large{}
}\textcolor{black}{{} we conjecture that using available asymmetry data
in low $x$ region may contribute to this difference. }

\subsection{$g_{1}$ Structure Functions}

\begin{figure}
\includegraphics[clip,width=0.45\textwidth]{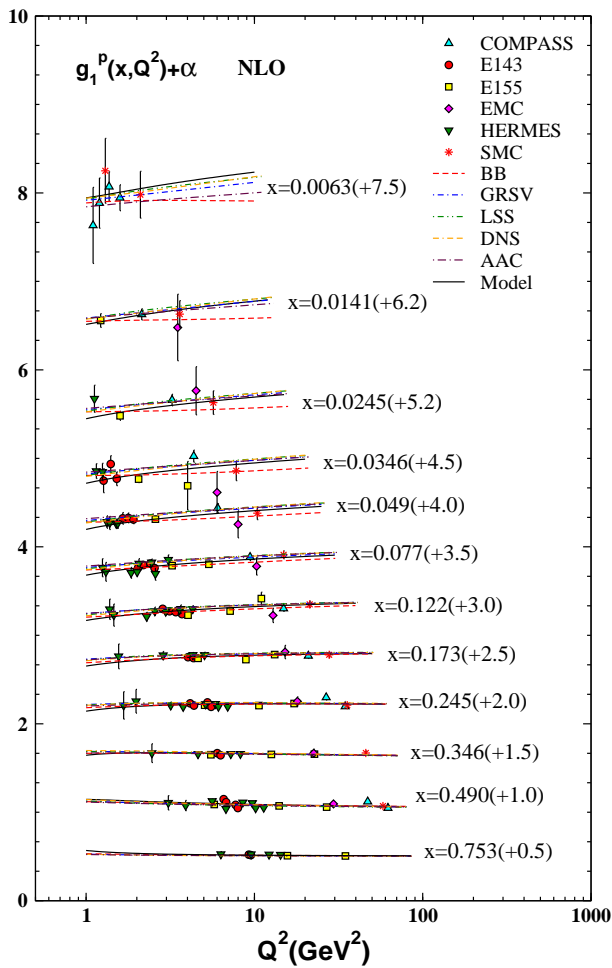}

\caption{{\small The polarized structure function $g_{1}^{p}$ as function
of $Q^{2}$ in intervals of $x$. The error bars shown are the statistical
and systematic uncertainties added in quadrature. Our fit is the solid
curve. The values of the shift $\alpha$ are given in parentheses.
Also shown are the results of BB (dashed) \citep{Bluemlein:2002be},
GRSV (dashed-dotted) \citep{Gluck:2000dy}, LSS (dashed-dotted-dotted)
\citep{Leader:2005ci}, DNS (dashed-dashed-dotted) \citep{deFlorian:2005mw}
and AAC (long dashed-dotted) \citep{Goto:1999by}. \label{fig:g1compare}}}

\end{figure}

\begin{figure}
\includegraphics[clip,width=0.4\textwidth]{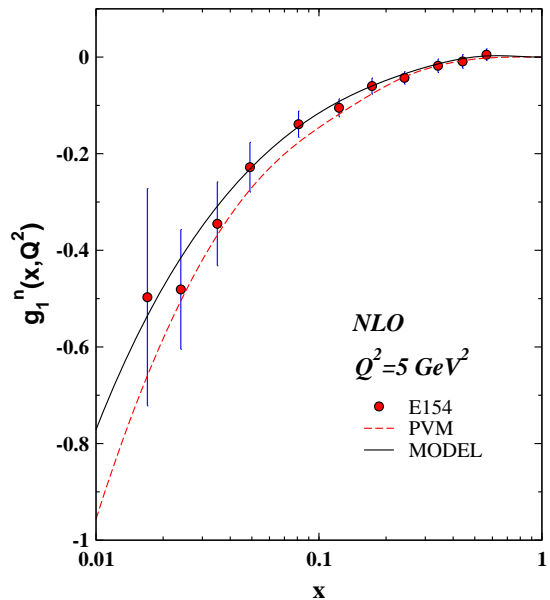}

\caption{{\small The polarized structure function $xg_{1}^{n}$ as function
of $x$ and for a fixed value of $Q^{2}=5$ GeV$^{2}$. The present
fit is the solid curve. Also shown are the results of AK \citep{Atashbar Tehrani:2007be}
(dashed) according to polarized valon model (PVM). \label{fig:g1n}}}

\end{figure}

\begin{figure*}
\includegraphics[clip,width=0.9\textwidth]{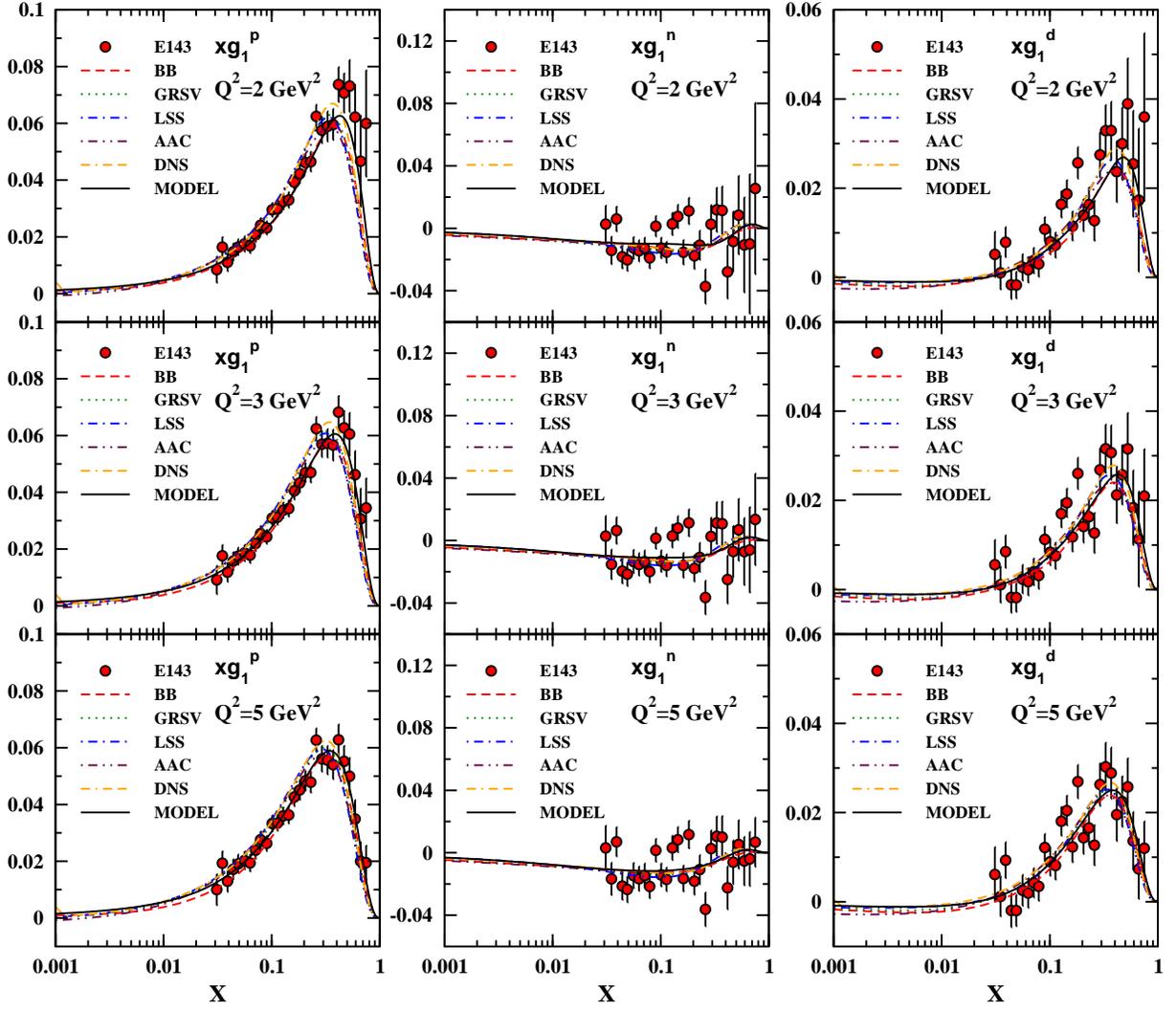}

\caption{{\small The polarized structure function $xg_{1}^{p}$, $xg_{1}^{n}$
and $xg_{1}^{d}$ as a function of $x$ for selected values of $Q^{2}$.
The data are well described by the fit (solid curve). Also shown are
the QCD NLO curves obtained by BB (dashed) \citep{Bluemlein:2002be},
GRSV (dotted) \citep{Gluck:2000dy}, LSS (dashed-dotted) \citep{Leader:2005ci},
AAC (dashed-dotted-dotted) \citep{Goto:1999by} and DNS (dashed-dashed-dotted)
\citep{deFlorian:2005mw}. \label{fig:g1pnd}}}

\end{figure*}

\begin{figure}
\includegraphics[clip,width=0.48\textwidth]{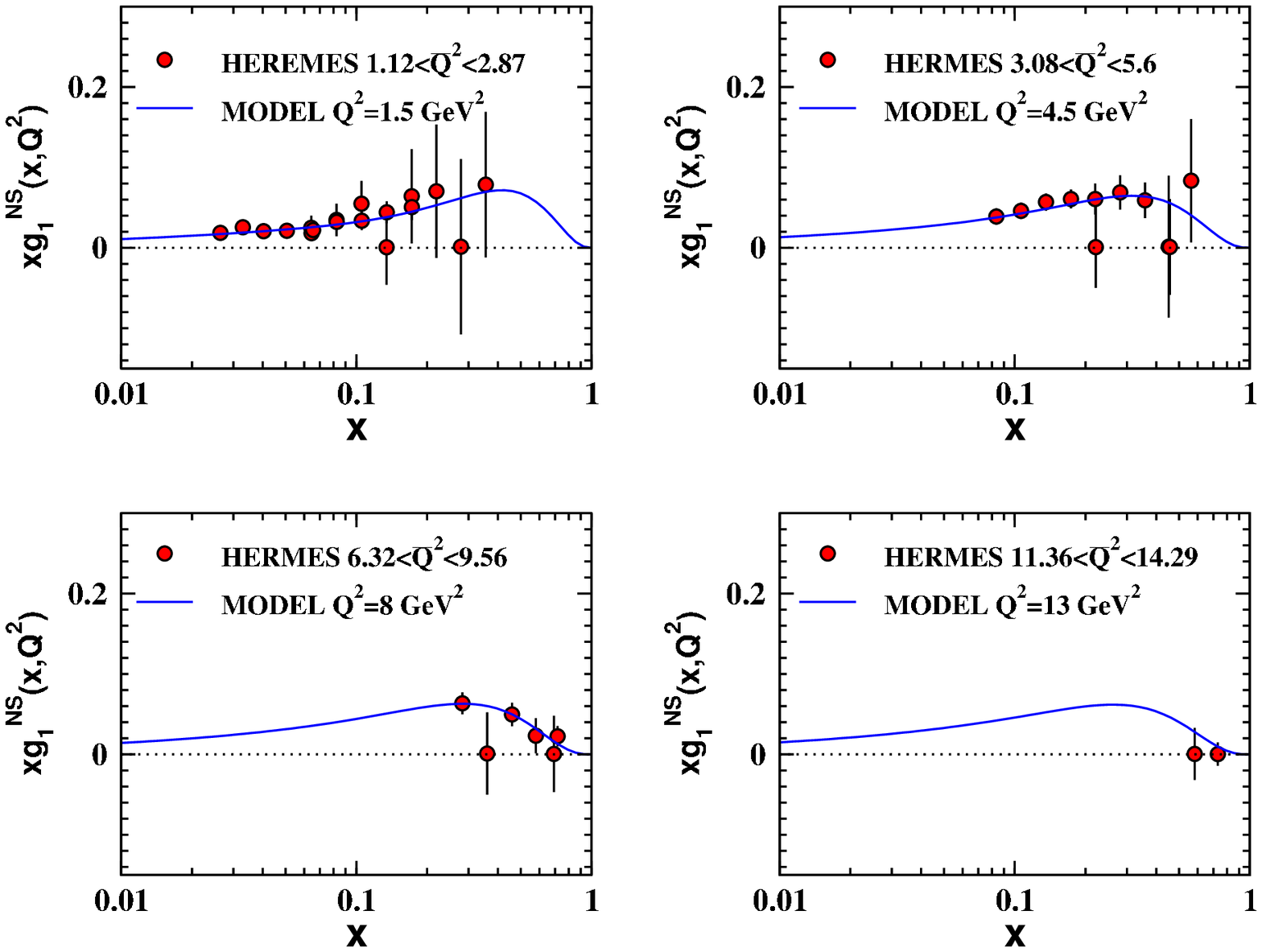}

\caption{{\small The non-singlet polarized structure function $xg_{1}^{NS}$
as function of $x$. \label{fig:g1ns}}}

\end{figure}

Figure~\ref{fig:g1compare} displays results for the polarized structure
function $xg_{1}^{p}$. For comparison, we display the results obtained
by (Blumlein, Bottcher) BB \citep{Bluemlein:2002be}, (Gluck, Reya,
Stratmann, Vogelsang) GRSV \citep{Gluck:2000dy}, (Leader, Sidorov,
Stamenov) LSS \citep{Leader:2005ci}, (de Florian, Navarro, Sassot)
DNS \citep{deFlorian:2005mw} and (Asymmetry Analysis Collaboration)
AAC \citep{Goto:1999by}. There is some spread in the analyses at
low values of $x$; however, the data are generally well described
within errors. As in the unpolarized case, the presence of scaling
violations result a slope that varies with changing $x$ values; this
is evident in Figure~\ref{fig:g1compare} where we observe the $Q^{2}$
dependence of the structure function $g_{1}(x,Q^{2})$.

Given the polarized proton PDFs, we can use isospin symmetry to obtain
the corresponding neutron structure functions. In Figure~\ref{fig:g1n},
we plot the neutron polarized structure function $xg_{1}^{n}$. We
also display the NLO QCD curves obtained by Ref.~\citep{Atashbar Tehrani:2007be}
in the polarized valon model (PVM).

We can relate the deuteron structure function to that of the proton
and neutron via: \begin{equation}
\textbf{M}[g_{1}^{d},N]=\frac{1}{2}\left.(1-\frac{3}{2}~\omega_{D}\right.)\left(\textbf{M}[g_{1}^{p},N]+\textbf{M}[g_{1}^{n},N]\right){\color{blue}\;,}\label{eq:g1d}\end{equation}
 where $\omega_{D}=0.05\pm0.01$ is the $D$-state wave probability
for the deuteron \citep{OMEGD}. In Figure~\ref{fig:g1pnd} we present
our results for the structure functions $xg_{1}^{p}(x,Q^{2})$, $xg_{1}^{n}(x,Q^{2})$
and $xg_{1}^{d}(x,Q^{2})$, and this compares favorably with the results
of the BB \citep{Bluemlein:2002be}, GRSV \citep{Gluck:2000dy}, LSS
\citep{Leader:2005ci}, DNS \citep{deFlorian:2005mw} and AAC \citep{Goto:1999by}
analyzes.

The non-singlet spin structure function $xg_{1}^{NS}(x,Q^{2})$ is
defined as \citep{HERMpd}\begin{eqnarray}
xg_{1}^{NS}(x,Q^{2}) & \equiv & xg_{1}^{p}(x,Q^{2})-xg_{1}^{n}(x,Q^{2})\nonumber \\
 & = & 2[xg_{1}^{p}(x,Q^{2})-\frac{xg_{1}^{d}(x,Q^{2})}{1-\frac{3}{2}\omega_{D}}]~.\label{eq:g1ns}\end{eqnarray}
 This is displayed in Figure~\ref{fig:g1ns}, and we compare with
the HERMES data \citep{HERMpd} for various $Q^{2}$ bins. In the
second line of Eq.~(\ref{eq:g1ns}) we have related the structure
function of the deuteron using isospin symmetry and the relation of
Eq.~(\ref{eq:g1d}).

\subsection{$g_{2}$ Structure Function}

\begin{figure}
\includegraphics[clip,width=0.4\textwidth]{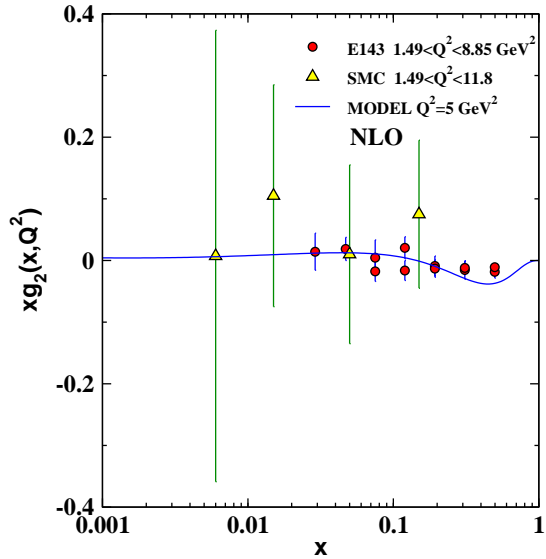}

\caption{{\small The polarized structure function $xg_{2}$ as
function of $x$ for $Q^{2}=5$ GeV$^2$. \label{fig:xg2}}}
\end{figure}

We can now extract the structure function $xg_{2}$ via the Wandzura-Wilczek
relation \citep{Wandzura:1977qf,Piccione:1997zh}:

\begin{equation}
g_{2}(x,Q^{2})=-g_{1}^{p}(x,Q^{2})+\int_{x}^{1}\frac{dy}{y}g_{1}^{p}(y,Q^{2})~.\label{eq:xg2}\end{equation}
 This relation remains valid in the presence of target mass corrections.
In Figure~\ref{fig:xg2} we show our result for $xg_{2}$ and we
compare it with the experimental data from E143 \citep{E143pd} and
SMC \citep{SMCpd}.

\subsection{First moment of $g_{1}$ structure functions}

\begin{table}
\begin{tabular}{ccccc}
\hline
Q$^{2}$  & 2 GeV$^{2}$  & 3 GeV$^{2}$  & 5 GeV$^{2}$  & 10 GeV$^{2}$ \tabularnewline
\hline
\hline
$\Delta u_{v}$  & 0.928864  & 0.928310  & 0.927794  & 0.927288 \tabularnewline
$\Delta d_{v}$  & -0.342318  & -0.342114  & -0.341924  & -0.341738\tabularnewline
$\Delta\overline{q}$  & -0.053400  & -0.053893  & -0.054379  & -0.054789\tabularnewline
$\Delta g$  & 0.143610  & 0.191313  & 0.248845  & 0.323886\tabularnewline
$\Gamma_{1}^{p}$  & 0.128291  & 0.131199  & 0.133822  & 0.136303\tabularnewline
$\Gamma_{1}^{n}$  & -0.050972  & -0.052416  & -0.053735  & -0.055000 \tabularnewline
$\Gamma_{1}^{d}$  & 0.035296  & 0.035965  & 0.036559  & 0.037115\tabularnewline
\hline
\end{tabular}

\caption{{\small The first moments of polarized parton distributions, $\Delta u_{v}$,
$\Delta d_{v}$, $\Delta\overline{q}$, $\Delta g$ and polarized
structure functions $\Gamma_{1}^{p},$ $\Gamma_{1}^{n}$, $\Gamma_{1}^{d}$
in NLO in the $\overline{{\rm MS}}$--scheme for some different values
of $Q^{2}$. \label{tab:firstMomQ}}}

\end{table}

\begin{table}
\begin{tabular}{|l|r|r|r|r|}
\hline
 & Model  & BB \citep{Blumlein:2010rn}  & GRSV~\citep{Gluck:2000dy}  & AAC~\citep{Goto:1999by}\tabularnewline
\hline
\hline
$\Delta u_{v}$  & $0.928$  & $0.928$  & 0.9206  & 0.9278 \tabularnewline
$\Delta d_{v}$  & $-0.342$  & $-0.342$  & --0.3409  & --0.3416 \tabularnewline
$\Delta u$  & $0.874$  & $0.866$  & 0.8593  & 0.8399 \tabularnewline
$\Delta d$  & $-0.396$  & $-0.404$  & --0.4043  & --0.4295 \tabularnewline
$\Delta\overline{q}$  & $-0.054$  & $-0.062$  & --0.0625  & --0.0879 \tabularnewline
$\Delta g$  & $0.224$  & $0.462$  & 0.6828  & 0.8076 \tabularnewline
\hline
\end{tabular}

\caption{{\small Comparison of the first moments of the polarized parton densities
in NLO in the $\overline{{\rm MS}}$--scheme at $Q^{2}=4$ }GeV$^{2}${\small{}
for different sets of recent parton parameterizations. The second
column (Model) contains the first moments which is obtained from our
new parametrization based on the Jacobi polynomials expansion method.
The BB \citep{Blumlein:2010rn}, GRSV \citep{Gluck:2000dy} and AAC
\citep{Goto:1999by} results are also shown. \label{tab:firstMom}}}

\end{table}

We next use the polarized PDFs to compute the first moments, and compare
with other recent analyzes. We can obtain the first moment of $g_{1}^{p}$
by \begin{equation}
\Gamma_{1}^{p}(Q^{2})\equiv\int_{0}^{1}dxg_{1}^{p}(x,Q^{2})\;.\end{equation}
 The results of our fit are presented in Table~\ref{tab:firstMomQ}
for selected values of $Q^{2}$, and these are compared with results
from the literature in Table~\ref{tab:firstMom}.

In the framework of QCD the spin of the proton can be expressed in
terms of the first moment of the total quark and gluon helicity distributions
and their orbital angular momentum, i.e. \begin{equation}
\frac{1}{2}=\frac{1}{2}\Delta\Sigma^{p}+\Delta g^{p}+L_{z}^{p}\;,\end{equation}
 where $L_{z}^{p}$ is the total orbital angular momentum of all quarks
and gluons. The contribution of $\frac{1}{2}\Delta\Sigma+\Delta g$
for typical value of $Q^{2}=4$ GeV$^{2}$ is around 0.355 in our
analysis. We can also compare this value in NLO with other recent
analysis. The reported value from the BB model \citep{Blumlein:2010rn}
is 0.569, the AAC model \citep{Goto:1999by} is 0.837 and the GRSV
model \citep{Gluck:2000dy} is 0.785, while the DSSV model \citep{deFlorian:2008mr}
is approximately 0.1. Since the values of $\frac{1}{2}\Delta\Sigma$
are comparable, we observe that the difference between the above reported
values must come from different gluon distributions.

\subsection{Strong Coupling Constant}
\begin{verse}
\begin{table}[t]
 \begin{tabular}{|c|c|c|c|}
\hline
$\alpha_{s}(M_{Z}^{2})$  & Order  & Reference  & Notes \tabularnewline
\hline
\hline
$0.1149\pm0.0015$  & NLO  &  & This analysis\tabularnewline
\hline
$0.1132_{-0.0095}^{+0.0056}$  & NLO  & \citep{Blumlein:2010rn}  & \tabularnewline
\hline
$0.1134_{-0.0021}^{+0.0019}$  & NNLO  & \citep{Blumlein:2006be}  & \tabularnewline
\hline
$0.1141\pm0.0036$  & NLO  & \citep{Atashbar
Tehrani:2007be}  & \tabularnewline
\hline
$0.1131\pm0.0019$  & NNLO  & \citep{Khorramian:2008yh}  & \tabularnewline
\hline
$0.1139\pm0.0020$  & NNNLO  & \citep{Khorramian:2009xz}  & \tabularnewline
\hline
$0.1141_{-0.0022}^{+0.0020}$  & NNNLO  & \citep{Blumlein:2006be}  & \tabularnewline
\hline
$0.1135\pm0.0014$  & NNLO  & \citep{ABKM}  & FFS \tabularnewline
\hline
$0.1129\pm0.0014$  & NNLO  & \citep{ABKM}  & BSMN \tabularnewline
\hline
$0.1124\pm0.0020$  & NNLO  & \citep{JR}  & dynamic approach \tabularnewline
\hline
$0.1158\pm0.0035$  & NNLO  & \citep{JR}  & standard approach \tabularnewline
\hline
$0.1171\pm0.0014$  & NNLO  & \citep{MSTW}  & \tabularnewline
\hline
$0.1147\pm0.0012$  & NNLO  & \citep{ABM}  & \tabularnewline
\hline
$0.1145\pm0.0042$  & NNLO  & \citep{H1ZEUS}  & (Preliminary) \tabularnewline
\hline
$0.1184\pm0.0007$  & ---  & \citep{Bethke:2009jm}  & World Average \tabularnewline
\hline
\end{tabular}

\caption{Comparison of $\alpha_{s}(M_{Z})$ values from the literature. \label{tab:as}}

\end{table}

\end{verse}
In this QCD analysis we extract $\alpha_{s}(Q_{0}^{2})$ at NLO and
obtain

\begin{equation}
\alpha_{s}(Q_{0}^{2})=0.381\pm0.017~.\end{equation}

\noindent Rescaling this to the $Z$ boson mass scale we find

\noindent \begin{equation}
\alpha_{s}(M_{Z}^{2})=0.1149\pm0.0015~.\label{eq:alphaMZ}\end{equation}
 The error given in above does not include the relative systematics
of the different classes of measurements. In Table~\ref{tab:as}
we provide a comparison of this value with other determinations from
the literature computed at NLO and higher orders, including the current
world average of $\alpha_{s}(M_{Z}^{2})=0.1184\pm0.0007$.

\subsection{Nuclear Polarized Structure Functions}

\begin{table}
\begin{tabular}{l|l>{\raggedright}p{0.1in}l}
\hline
\multicolumn{1}{c}{n} & \multicolumn{3}{c}{p}\tabularnewline
\hline
\fstrut $a^{n}=5.650556817$  & $a_{0}^{p}=0.0148376$  &  & $b_{0}^{p}=4.15388$\tabularnewline
$b^{n}=0.986818274$  & $a_{1}^{p}=-0.0189575$  &  & $b_{1}^{p}=-4.75525$\tabularnewline
$c^{n}=0.064446823$  & $a_{2}^{p}=0.0121792$  &  & $b_{2}^{p}=2.68417$\tabularnewline
$d^{n}=0.807650292$  & $a_{3}^{p}=-0.0040397$  &  & $b_{3}^{p}=-0.800306$\tabularnewline
 & $a_{4}^{p}=0.000540845$  &  & $b_{4}^{p}=0.101095$ \tabularnewline
\hline
\end{tabular}\caption{{\small Numerical coefficients for Eqs.~(\ref{eq:f3HeN},\ref{eq:f3HeP})
of $\Delta f_{^{3}He}^{n}(y)$ and $\Delta f_{^{3}He}^{p}(y)$ obtained
from Refs.~\citep{Bissey:2000ed,Bissey:2001cw,Afnan:2003vh}. \label{tab:npHe}}}

\end{table}

\begin{figure}
\includegraphics[clip,width=0.3\textwidth]{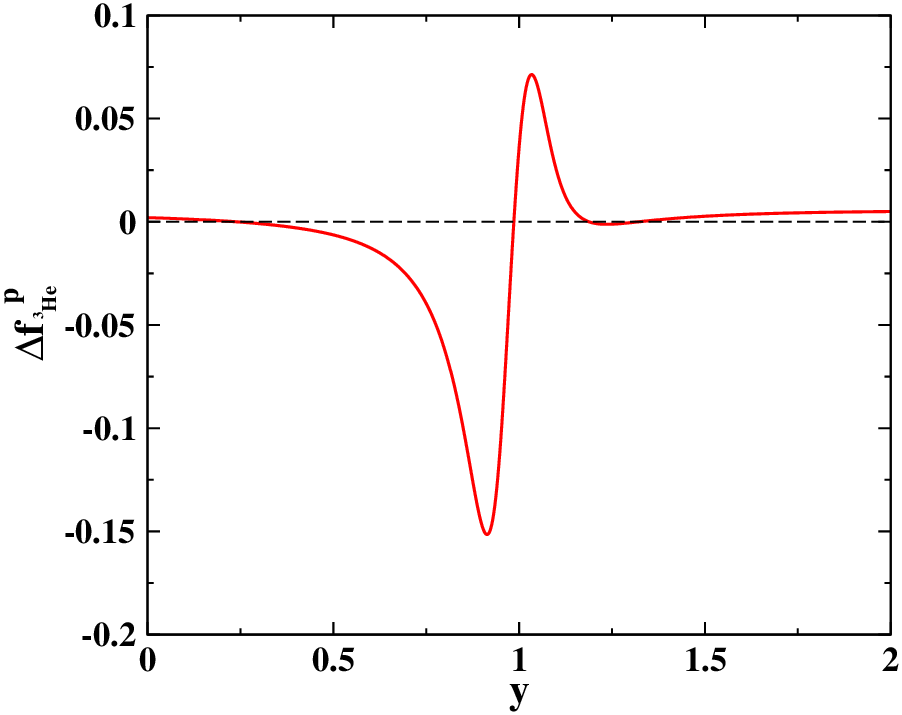}

\caption{{\small The polarized light cone distribution function for the proton
in the $^{3}He$, based on the results of Ref.~\citep{Bissey:2001cw,Bissey:2000ed,Afnan:2003vh}.
\label{fig:fpHe}}}

\end{figure}

\begin{figure}
\includegraphics[clip,width=0.3\textwidth]{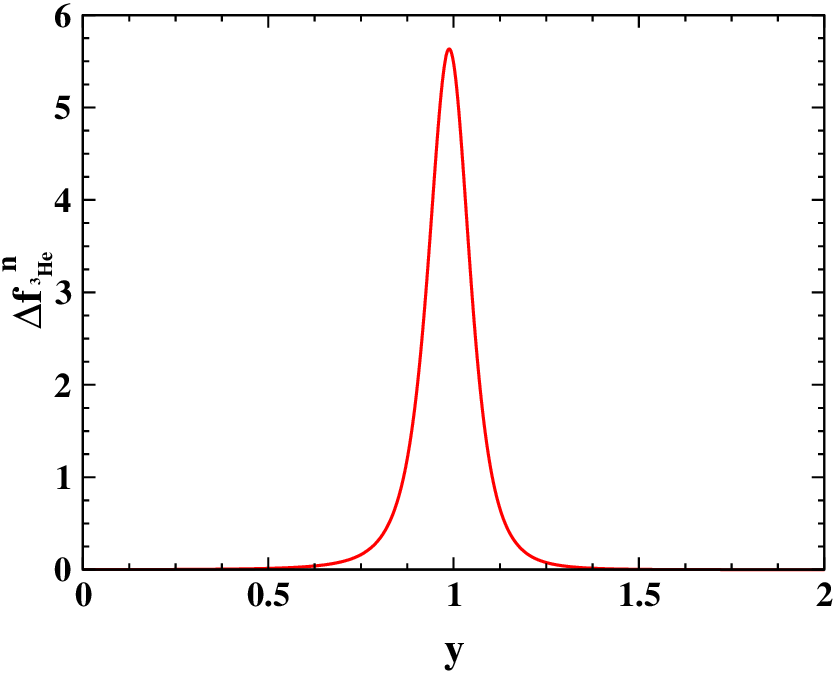}

\caption{{\small The polarized light cone distribution function for the neutron
in the $^{3}He$, based on the results of Ref.~\citep{Bissey:2001cw,Bissey:2000ed,Afnan:2003vh}.
\label{fig:fnHe}}}

\end{figure}

\begin{figure}
\includegraphics[clip,width=0.35\textwidth]{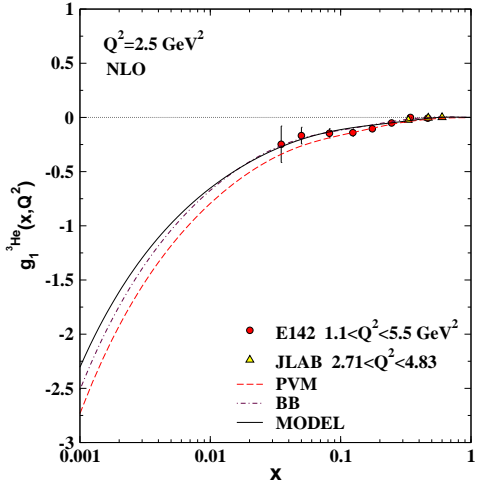}

\caption{{\small Analytical result for the polarized $^{3}$He structure function
}\textit{\small v.s.}{\small{} $x$ for fixed $Q^{2}=2.5$ GeV$^{2}$.
The current fit is the solid curve. Also shown are the QCD NLO curves
obtained by AK (dashed) \citep{Atashbar Tehrani:2007be} according
to polarized valon model (PVM) and BB (dashed-dotted) \citep{Bluemlein:2002be}.
\label{fig:g1he}}}

\end{figure}

\begin{figure}
\includegraphics[clip,width=0.35\textwidth]{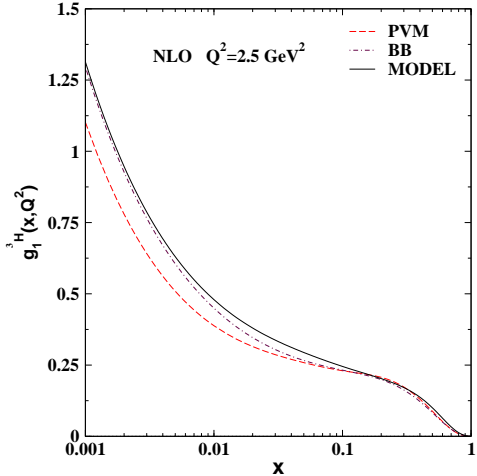}

\caption{{\small Analytical result for the polarized $^{3}$H structure function
}\textit{\small v.s.}{\small{} $x$ for fixed $Q^{2}=2.5$ GeV$^{2}$.
The current fit is the solid curve. Also shown are the QCD NLO curves
obtained by AK (dashed) \citep{Atashbar Tehrani:2007be} according
to polarized valon model (PVM) and BB (dashed-dotted) \citep{Bluemlein:2002be}
for comparison. \label{fig:g1h}}}

\end{figure}

Using the polarized PDF fit results, we examine the nucleon corrections
factors for ${^{3}He}$ and ${^{3}H}$. The polarized structure functions
$g_{1}^{^{3}He}$ and $g_{1}^{^{3}H}$ can be composed from the polarized
proton structure $g_{1}^{p}$ and the polarized neutron structure
$g_{1}^{n}$ as follows:

\begin{eqnarray}
g_{1}^{^{3}He}(x,Q^{2}) & = & \int_{x}^{3}\frac{dy}{y}\Delta f_{^{3}He}^{n}(y)g_{1}^{n}(\frac{x}{y},Q^{2})\nonumber \\
 & + & 2\int_{x}^{3}\frac{dy}{y}\Delta f_{^{3}He}^{p}(y)g_{1}^{p}(\frac{x}{y},Q^{2})\nonumber \\
 & - & 0.014[g_{1}^{p}(x,Q^{2})-4g_{1}^{n}(x,Q^{2})]\;,\label{eq:g1He}\end{eqnarray}
 \begin{eqnarray}
g_{1}^{^{3}H}(x,Q^{2}) & = & 2\int_{x}^{3}\frac{dy}{y}\Delta f_{^{3}H}^{n}(y)g_{1}^{n}(\frac{x}{y},Q^{2})\nonumber \\
 & + & \int_{x}^{3}\frac{dy}{y}\Delta f_{^{3}H}^{p}(y)g_{1}^{p}(\frac{x}{y},Q^{2})\nonumber \\
 & + & 0.014[g_{1}^{p}(x,Q^{2})-4g_{1}^{n}(x,Q^{2})]\;.\label{eq:g1H}\end{eqnarray}
 Here, $\Delta f_{^{3}He}^{N}(y)$ and $\Delta f_{^{3}H}^{N}(y)$
are the spin-dependent nucleon light-cone momentum distributions \citep{Yazdanpanah:2009zz,Bissey:2001cw}.
These functions parameterize the Fermi motion and the nucleon binding,
and are readily calculated using the ground-state wave functions of
$^{3}He$ and $^{3}H$. Note that the last term in above equations
is important only in the large-$x$ region.

If we utilize isospin symmetry, we can equate $\Delta f_{^{3}He}^{p}(y)$
to $\Delta f_{^{3}H}^{n}(y)$, and also $\Delta f_{^{3}He}^{n}(y)$
to $\Delta f_{^{3}H}^{p}(y)$; thus, we are left with only two independent
functions $\Delta f_{^{3}He}^{p}(y)$ and $\Delta f_{^{3}He}^{n}(y)$.
Using the results of Refs.~\citep{Bissey:2001cw,Bissey:2000ed,Afnan:2003vh},
we express these distributions as

\begin{equation}
\Delta f_{^{3}He}^{n}(y)=\frac{a^{n}e^{-\frac{0.5(1-d^{n})(-b^{n}+y)^{2}}{{(c^{n})}^{2}}}}{1+\frac{d^{n}(-b^{n}+y)^{2}}{{(c^{n})}^{2}}}~,\label{eq:f3HeN}\end{equation}
 \begin{equation}
\Delta f_{^{3}He}^{p}(y)=\frac{\sum_{i=0}^{4}a_{i}^{p}U_{i}(y)}{\sum_{i=0}^{4}b_{i}^{p}U_{i}(y)}~,\label{eq:f3HeP}\end{equation}
 where $U_{n}(y)$ is a Chebyshev polynomials of the second type.
The numerical coefficients of these equations are presented in Table~\ref{tab:npHe}.
We can then use Eqs.~(\ref{eq:g1He},\ref{eq:g1H}) to obtain the
polarized nucleon structure functions $g_{1}^{^{3}He}(x,Q^{2})$ and
$g_{1}^{^{3}H}(x,Q^{2})$.

To determine the $g_{1}^{^{3}He}$ and $g_{1}^{^{3}H}$ polarized
structure functions we need the polarized light-cone distribution
functions for proton and neutron in $^{3}He$, i.e. $\Delta f_{{^{3}}He}^{p}$
and $\Delta f_{{^{3}}He}^{n}$. In Figures~\ref{fig:fpHe} and \ref{fig:fnHe}
we present our results using the parametrization of Eqs.~(\ref{eq:f3HeN},\ref{eq:f3HeP})
which is based on the numerical results of Ref.~\citep{Afnan:2003vh}.

\textcolor{black}{In Figures~\ref{fig:g1he} and \ref{fig:g1h} we
show our results for the $g_{1}^{^{3}He}$ and $g_{1}^{^{3}H}$
polarized structure function, and compare with
BB~\citep{Bluemlein:2002be}, and the polarized valon model (PVM)
\citep{Atashbar Tehrani:2007be}. For the $g_{1}^{^{3}He}$
polarized structure function we see that our result coincides with
the BB fit for $x$ values down to $\sim10^{-2}$, and then falls
off more quickly at very small $x$ values. The polarized valon
model (PVM), while still a reasonable fit to the data, lies below
both of the other fits. For the $g_{1}^{^{3}H}$ polarized
structure function, our fit coincides with the BB fit at both
large and small $x$ values, but dips below it (closer to the PVM)
for intermediate $x$ values. The differences between these curves
come from the various data sets used, the constraints imposed, and
the form of the parameterization. For example, in the AK fit
\citep{Atashbar Tehrani:2007be}, only 257 experimental data points
were used as the neutron data were not included; in contrast, the
present analysis uses 379 points which does include the neutron
data. Furthermore, the AK fit used 15 free parameters while there
are only 9 free parameters in the present analysis. These
differences are reflected in the extractions of PPDFs, and a
comparison of these different analyses may be indicative of the
stability of the determined QCD parameters.}

\subsection{Bjorken Sum Rule}

We can also study the Bjorken sum rule \citep{Bjorken:1966jh} which
relates the difference of the first moments of the proton and neutron
spin structure functions to the axial vector coupling constant of
the neutron $\beta$-decay, \begin{equation}
\int_{0}^{1}[g_{1}^{p}(x,Q^{2})-g_{1}^{n}(x,Q^{2})]dx=\frac{1}{6}~g_{A}[1+{\it O}(\frac{\alpha_{s}}{\pi})]~,\label{eq:bsum}\end{equation}
 where $g_{A}=1.2670\pm0.0035$ \citep{PDG}, and the QCD radiative
corrections are denoted as ${\it O}(\frac{\alpha_{s}}{\pi})$. This
sum rule can be generalized for the $^{3}$He--$^{3}$H system as
follows: \begin{equation}
\int_{0}^{3}[g_{1}^{^{3}H}(x,Q^{2})-g_{1}^{^{3}He}(x,Q^{2})]dx=\frac{1}{6}~\tilde{g}_{A}[1+{\it O}(\frac{\alpha_{s}}{\pi})]~,\label{eq:bsumnew}\end{equation}
 where $\tilde{g}_{A}$ is the axial vector coupling constant of the
Triton $\beta$-decay, with $\tilde{g}_{A}=1.211\pm0.002$ \citep{Budick:1991zb}.
Taking the ratio of the Eqs.~(\ref{eq:bsum}) and (\ref{eq:bsumnew}),
we find \begin{equation}
\frac{\int_{0}^{3}[g_{1}^{^{3}H}(x,Q^{2})-g_{1}^{^{3}He}(x,Q^{2})]dx}{\int_{0}^{1}[g_{1}^{p}(x,Q^{2})-g_{1}^{n}(x,Q^{2})]dx}=\frac{\tilde{g}_{A}}{g_{A}}~.\label{eq:ratio}\end{equation}
 Given $g_{A}$ and ${\tilde{g}_{A}}$, we compute the above ratio
to be 0.956 \citep{Bissey:2001cw}. Note that the QCD radiative corrections
are expected to cancel exactly in above equation. Using the Bjorken
sum rules of Eqs.~(\ref{eq:bsum},\ref{eq:bsumnew}), we obtain the
value 0.924 for the ratio of Eq.~(\ref{eq:ratio}).

\section{Conclusions}

We have presented a fit to the polarized lepton-DIS data on nuclei
at NLO QCD using the Jacobi polynomial method. Having extracted the
polarized PDFs, we compute various nuclear structure functions $(g_{1},g_{2})$
and Bjorken sum rule. In general, we find good agreement with the
experimental data, and our results are in accord with other determinations
from the literature; collectively, this demonstrates progress of the
field toward a detailed description of the spin structure of the nucleon.

Having demonstrated the compatibility of the Jacobi polynomial method
with other approaches in the literature, this study can serve as a
foundation for addressing issues of polarized scattering processes
from a complementary perspective. In particular, the Jacobi polynomial
method offers the opportunity to examine efficiencies of different
methods, and this work is in progress.

\section*{Acknowledgments}

We thank S. Kumano and M. Miyama of the ACC Collaboration for allowing
us to use their interpolation routines. A.N.K. thanks Johannes Blümlein
for useful discussions, and is grateful to A. L. Kataev for the suggestion
of the Jacobi polynomial method, the CERN TH-PH division for the hospitality
where a portion of this work was performed, and Semnan University
for financial support. We acknowledge financial support of the the
School of Particles and Accelerators, Institute for Research in Fundamental
Sciences (IPM). This work was partially supported by the U.S.\ Department
of Energy under grant DE-FG02-04ER41299, and the Lightner-Sams Foundation.

\section*{Appendix: FORTRAN-code }

A \texttt{FORTRAN} package containing our $g_{1}(x,Q^{2})$ polarized
structure functions for $\{p,n,d,NS,{}^{3}He,{}^{3}H\}$ and $xg_{2}^{p}(x,Q^{2})$,
as well as the polarized parton densities $\{u_{v,}d_{v},g,\bar{q}\}$.
$x\delta u_{v}(x,Q^{2})$, $x\delta d_{v}(x,Q^{2})$, $x\delta g(x,Q^{2})$
and $x\delta\bar{q}(x,Q^{2})$ at NLO in the $\overline{{\rm MS}}$--scheme
can be found in \texttt{http://particles.ipm.ir/links/QCD.htm} \quad{}or
obtained via e-mail from the authors. These functions are interpolated
using cubic splines in $Q^{2}$ and a linear interpolation in $\log\,(Q^{2})$.The
package includes an example program to illustrate the use of the routines.

\end{document}